\documentclass[useAMS]{mn2e}

\usepackage{amssymb,amsmath,psfig,times}
\def\gsim{ \lower .75ex \hbox{$\sim$} \llap{\raise .27ex \hbox{$>$}} }
\def\lsim{ \lower .75ex\hbox{$\sim$} \llap{\raise .27ex \hbox{$<$}} }

\def\sc{Schwarzschild}
\def\beq{\begin{equation}}
\def\eeq{\end{equation}}

\title[The blazar S5 0014+813]
{The blazar S5 0014+813: a real or apparent monster?}  
\author[G. Ghisellini et al.]
{G. Ghisellini$^1$\thanks{Email:
gabriele.ghisellini@brera.inaf.it}, L. Foschini$^1$, M. Volonteri$^2$, 
G. Ghirlanda$^1$, F. Haardt$^3$, D. Burlon$^4$,
\newauthor F. Tavecchio$^1$
\\
$^1$INAF -- Osservatorio Astronomico di Brera, Via Bianchi 46, I--23807 Merate, Italy\\
$^2$Astronomy Department, University of Michigan, Ann Arbor, MI 48109 \\
$^3$Universit\'a dell'Insubria, Dipartimento di Scienze Chimiche, Fisiche e 
Matematiche, Via Valleggio 11, 22100 Como, Italy;\\
$^4$Max Planck Institut f\"ur extraterrestrische Physik, Giessenbachstrasse 1, 85748 Garching, Germany \\
}

\begin{document}  

\maketitle

\begin{abstract}
A strong hard X--ray luminosity from a blazar flags the presence
of a very powerful jet. 
If the jet power is in turn related to the mass accretion rate, the most
luminous hard X--ray blazars should pinpoint the largest accretion rates,
and thus the largest black hole masses. 
These ideas are confirmed by the {\it Swift} satellite observations of 
the blazar S5 0014+813, at the redshift $z=3.366$. 
{\it Swift} detected this source
with all its three instruments, from the optical to the hard X--rays.
Through the construction of its spectral energy distribution we are confident that
its optical--UV emission is thermal in origin.
Associating it to the emission of a standard optically thick geometrically thin
accretion disk, we find a black hole mass
$M\sim 4\times 10^{10} M_\odot$, radiating at 40\% the Eddington value.
The derived mass is among the largest ever found. 
Super--Eddington slim disks or thick disks with the presence of a collimating funnel
can in principle reduce the black hole mass estimate, but
tend to produce spectra bluer than observed.
%
\end{abstract}

\section{Introduction}

Most of the estimates of the black hole mass in quasars
make use of the width of broad emission lines in their
spectra and an empirical relation between the luminosity of the 
ionising continuum and the size of the broad line region (BLR;
Kaspi et al. 2007, Bentz et al. 2006, 2009).
This, plus the assumption of virial velocities, allows to estimate
the black hole mass (albeit with some uncertainties).
In recent years the Sloan Digital Sky Survey (SDSS) of galaxies and quasars is  providing 
the largest quasars samples suitable to study the 
black hole mass of quasars, and how their corresponding  mass function 
evolves with redshift.
We here use a different (and ``more ancient")
method to estimate the black hole mass,
by directly interpreting the optical--UV flux of a
source as the emission produced by a standard accretion disk, namely
a Shakura--Sunjaev (1973)
multi--colour disk, emitting as a black--body at each annulus.

We do this exercise for carefully selected sources: blazars
that are very luminous in hard X--rays, above 10 keV.
The rationale for this choice is the following.
We know that the spectral energy distribution (SED) of blazars
is characterised by two broad humps, whose peak frequencies (in $\nu F_\nu$)
are a function of the observed bolometric luminosity of the blazars
(the so called ``blazar sequence"; Fossati et al. 1998; Ghisellini et al. 1998).
Larger powers correspond to smaller peak frequencies and more dominance
of the high energy peak to the low energy one.
According to this scenario, the most powerful blazars should have their
peaks in the far IR and in the $\sim$1 MeV bands.
The latter peak should also carry the bulk of the electromagnetic output.
This has two important consequences: i) the hard X--ray luminosity,
being close to the peak, is large; ii) the non--thermal (synchrotron)
radiation of the first peak (being in the far IR) 
does not hide the accretion disk radiation peaking at optical--UV frequencies.
Therefore in these kind of blazars we can study the radiation from the
accretion disk directly (see e.g. 
Landt et al. 2008; 
Maraschi et al. 2008; Sambruna et al. 2007).
The last and important steps link the observed luminosity
to the power carried by the jet (in bulk motion of particles and fields)
and the jet power to the mass accretion rate.
The latter point, investigated by Rawlings \& Saunders back in 1991, has
been since then confirmed by other groups and using different methods
(e.g. Celotti et al. 1997; 
Cavaliere \& D'Elia 2002; 
Maraschi \& Tavecchio 2003; 
Sambruna et al. 2006;
Allen et al. 2006;
Celotti \& Ghisellini 2008; 
Ghisellini \& Tavecchio 2008, 2009;
Kataoka et al. 2008).

From the above it is clear that having a sample of distant FSRQs detected
in hard X--rays allows to pinpoint the most accreting systems, and
thus large black hole masses and large accretion rates.
The recently published list of blazars detected by the 
Burst Alert Telescope (BAT) onboard the {\it Swift} satellite
(Ajello et al. 2009, hereafter A09) has provided the first well constructed sample
suitable for our aim.
It contains 26 Flat Spectrum Radio Quasars (FSRQs) and 12 BL Lacs
detected during three years of survey in the 15--55 keV range.
One of those, S5 0014+813 ($z=3.366$),
is exceptionally bright in the optical band. 
Adopting a cosmology with $\Omega_{\rm M}=0.3$ and 
$\Omega_\Lambda=h_0=0.7$ we have, in the optical, 
$\nu L_\nu\sim 10^{48}$ erg s$^{-1}$.
This is the source discussed in this paper, with the aim
to find the mass of its black hole and the corresponding accretion rate.

\section{The blazar S5 0014+813}

This FSRQ was discovered in the radio band by 
Kuhr et al. (1981), and it was soon noted as 
exceptionally luminous in the optical 
(Kuhr et al. 1983).
The spectrum taken by Sargent et al. (1989) had a (extinction corrected)
slope of $\alpha=0.8$ [$F(\nu)\propto \nu^{-\alpha}$] longward of the 
Ly$\alpha$ line of equivalent  width of 158 \AA.
Fried (1992) find no excess of foreground
galaxies in the direction
of the blazar, so disfavouring the hypothesis of gravitational lensing.
It is not polarised in the optical (Kuhr et al. 1983), and showed very mild
optical variability ($\max \Delta m\sim 0.15$ in 9 years;
Kaspi et al. 2007).
At a Galactic latitude is $18.8^\circ$, it suffers 
from a non negligible optical extinction.
Schlegel et al. (1998) lists $E(B-V)=0.19$, corresponding to  $A_V=0.62$.


This blazars was discussed in detail by Bechtold et al. (1994), that also
analysed {\it ROSAT} data and showed the overall SED.
These authors estimated the black hole mass on the basis
of the optical--UV luminosity, that was however severely underestimated.


The results of the  {\it XMM--Newton} observations on 23 Aug. 2001 
are presented by  Page et al. (2005). 
An absorbed power--law of photon index  $\Gamma=1.61\pm 0.02$
best fitted the data, with a column
$N_{\rm H}^{\rm host}=(1.8\pm 0.19)\times 10^{22}$ cm$^{-2}$ 
located at the redshift of the source, 
in addition to the Galactic one.
The [0.3--10 keV] flux was $F_X =5.6\times 10^{-12}$ erg cm$^{-2}$ s$^{-1}$,
corresponding to $L_X \sim 6\times 10^{47}$ erg s$^{-1}$.
Observed with VLBI, it showed no superluminal motion (Piner et al. 2007). 


\section{Swift observations and analysis}

{\it Swift} observed S5 0014+813 in January 2007. 
We analyzed these data with the 
most recent software \texttt{Swift\_Rel3.2} released as part of 
the \texttt{Heasoft v. 6.6.2}.
The calibration database is that updated to April 10, 2009. 


The XRT data were processed with the standard procedures ({\texttt{XRTPIPELINE v.0.12.2}). 
Source 
events were extracted in a circular region of  aperture $\sim 47''$, and background was estimated 
in a same sized circular region far from the source. Response matrices were created through the 
\texttt{xrtmkarf} task. 
We analysed the single observations separately and also summed them together. 
Each spectrum was analysed through XSPEC
with an absorbed power--law with a fixed 
Galactic column density ($N^{\rm Gal}_{\rm H}=1.32\times 10^{21}$ cm$^{-2}$ 
from Kalberla et al. 2005) 
The computed errors represent the 
90\% confidence interval on the spectral parameters.
The best fit photon index of the summed spectrum was $\Gamma=1.36\pm0.11$
with a $\chi^2=16$ for 19 degrees of freedom
and an observed (de--absorbed) flux $F_X=5.3\times 10^{-12}$ 
($F_X=6.1\times 10^{-12}$) erg cm$^{-2}$ s$^{-1}$.
All quantities are calculated in the [0.2--10 keV] band.

We have also re--analysed the {\it XMM--Newton} observation
using the {\texttt{SAS software (v. 9.0.0)}}. 
After screening the data for high background phases, we have
a net exposure of 11 ks.
%
%
Fitting the 0.2--10 keV energy range with a power law 
we obtain $\Gamma = 1.44\pm 0.05$, with   
$N_{\rm H}^{\rm host}=6\times 10^{21}$ cm$^{-2}$ 
(in addition to $N^{\rm Gal}_{\rm H}$), 
harder than obtained by Page et al. (2005).
This is likely due to our better screening and  the 
improved calibrations.
%
%
A broken power law (with $N_{\rm H}$ 
fixed to $N^{\rm Gal}_{\rm H}$) better fits the data (at the 99.99\% level), 
with a break at $E_{\rm b}=1.0\pm0.2$ keV and slopes $\Gamma_1=1.1\pm0.2$ and
$\Gamma_2=1.43\pm0.04$ below and above $E_{\rm b}$, respectively (see Fig. \ref{f1}).


UVOT (Roming et al. 2005) source counts were extracted from 
a circular region $5''-$sized centred on the source position, 
while the background was extracted from 
a larger circular nearby source--free region.
Data were integrated with the \texttt{uvotimsum} task and then 
analysed by using the  \texttt{uvotsource} task.

The observed magnitudes have been dereddened according to the formulae 
by Cardelli et al. (1989) and converted into fluxes by using standard 
formulae and zero points from Poole et al. (2008).
The source was detected in $V$, $B$ and $U$ (observed magnitudes
$V=16.43\pm 0.05$; $B=17.57\pm 0.06$ and $U=18.2\pm 0.1$, not corrected
by extinction), while only upper limits were obtained in the remaining
three filters ($UVW1>19.4$; $UVM2>19.6$ and $UVW2>20.0$, 3$\sigma$ limits).

\section{The spectral energy distribution}

Fig. \ref{f1} shows the SED of S5 0014+813.
The BAT data correspond to the average flux of the
3 years survey, while the UVOT and XRT data are the sum of three
pointed observations.
The grey empty symbols are archival data, while filled grey symbols 
and thick (optical) segments are from Bechtold et al. (1994).
The magenta points are from IRAS (Moshir et al. 1991) and 2MASS (Cutri et al. 2003).
The solid lines correspond to our modelling (see below).
This source is not in the list of blazars
detected in the first three months of {\it Fermi} (Abdo et al. 2009).
We have estimated the corresponding upper limit 
shown by the arrow. 
There must be a  peak between the BAT and the {\it Fermi} energy range.
There is a slight mismatch between the level of the BAT flux and 
the extrapolation from the XRT spectrum, while the {\it XMM--Newton} data well agree
with the XRT ones.
The BAT/XRT mismatch can be easily accounted for by considering that all blazars are
very variable sources (despite the XRT/{\it XMM--Newton} coincidental resemblance)
and the BAT flux is a 3--years average, while the shown XRT
spectrum is the sum of three observations within three days.

\begin{figure}
\vskip -0.5 cm
\hskip -0.3 cm
\psfig{figure=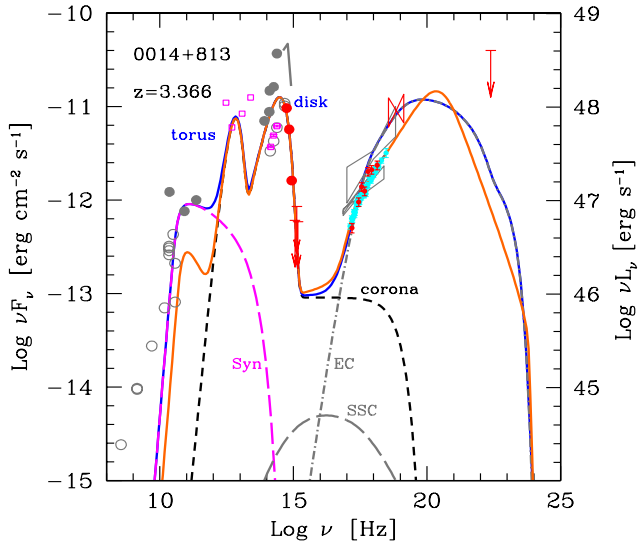,width=9cm,height=8.5cm}
\vskip -0.6 cm
\caption{
SED of S5 0014+813 together with the fitting models,
with parameters listed in Tab. \ref{para}.
UVOT, XRT and BAT data are indicated
by red symbols,
while archival data (form NED) are in light grey.
The magenta square symbols are IRAS and 2MASS data points.
The dotted line is the emission from the IR torus, 
the accretion disk and its X--ray corona. 
The blue and orange lines are the sum of all components.
The two models differ mainly for the location of the dissipation
region of the jet: outside (blue) or inside (orange)
the BLR.
}
\label{f1}
\end{figure}

The optical--UV spectrum is dominated by a a narrow bump,
that we interpret as the emission produced by the accretion disk.
This is substantiated by three facts:
i) the broad emission lines of the source are well visible (e.g.
Sargent et al. 1989; Bechtold et al. 1994; Osmer, Porter \& Green 1994) 
and are therefore not swamped by the non--thermal
continuum; ii) there is a general consensus to interpret the overall 
non--thermal SED of blazars, from the far IR to $\gamma$--rays,
as due to a single population of electrons.
If so, the relatively steep emission above the high energy peak
is made by electrons that emit, by synchrotron, a correspondingly steep
spectrum in the IR--optical--UV band.
This leaves the accretion disk component unhidden.
iii) A synchrotron and inverse Compton model that reproduces
the observed fluxes in the UVOT optical--UV and $\sim$MeV bands
fails to reproduce the very hard optical spectrum [that has a slope
$F(\nu)\propto \nu^0$] (although the non--simultaneity of the 
IR--optical data leaves some uncertainties).

Then we model the optical--UV flux by a standard 
multi--colour accretion disk, with a temperature profile 
given by (see e.g. Frank King \& Raine 2002):
\begin{equation}
T^4 \, =\, {  3 R_{\rm S}  L_{\rm d }  \over 16 \pi\eta\sigma_{\rm MB} R^3 }  
\left[ 1- \left( {3 R_{\rm S} \over  R}\right)^{1/2} \right]   
\end{equation}
where $L_{\rm d}=\eta \dot M c^2$ is the bolometric disk luminosity,
$R_{\rm S}$ is the \sc\ radius and $\sigma_{\rm MB}$ is the Maxwell--Boltzmann constant.
The maximum temperature (and hence the peak of the disk $\nu F_\nu$ spectrum)
occurs at $R\sim 5 R_{\rm S}$ and scales as
$T_{\rm max}\propto (L_{\rm d}/L_{\rm Edd})^{1/4}M^{-1/4}$.
The total optical--UV flux gives $L_{\rm d}$ 
[that of course scales as $(L_{\rm d}/L_{\rm Edd})\, M$].
Therefore we can derive both the black hole mass and the accretion rate.

The results of changing both are shown in Fig. \ref{f2},
where we show the accretion disk luminosity for 
a black hole of 10, 20 and 40 billion solar masses accreting at the Eddington 
or at the 40\% of the Eddington rate.
With $M=4\times 10^{10} M_\odot$ we can reproduce reasonably well both the
current state observed by UVOT and the old data discussed by Bechtold et al.
(1994)\footnote{Note that Bechtold et al. used a cosmology with 
$H_0=100$ and $q_0=0.5$,
giving a smaller distance than used here.}.
The two states would then differ because of the different accretion rate.

\begin{figure}
\vskip -0.5 cm
\hskip -0.3 cm
\psfig{figure=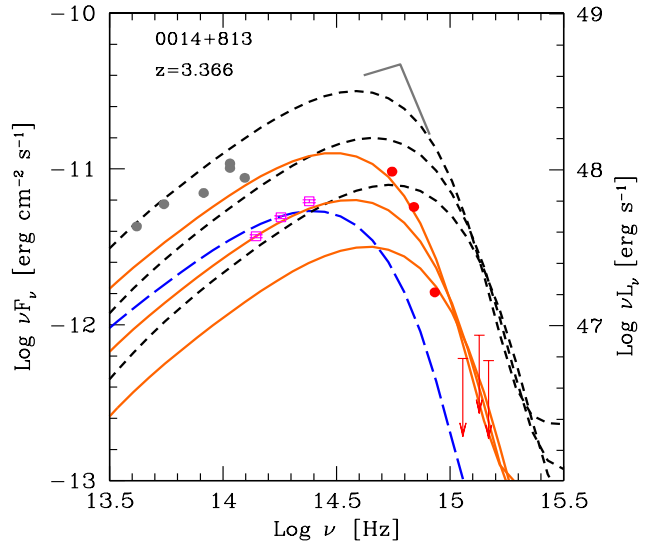,width=9cm,height=8.5cm}
\vskip -0.6 cm
\caption{
Zoom of the SED of S5 0014+813,
The three dashed lines correspond to $L_{\rm d}=L_{\rm Edd}$
for three black hole masses (40, 20 and 10 billion solar masses,
from top to bottom).
The three solid lines correspond to $L_{\rm d}=0.4 L_{\rm Edd}$
for the same three black hole masses.
The blue line correspond to $M=4\times 10^{10}$ and
$L_{\rm d}=0.17 L_{\rm Edd}$.
Note that $M=4\times 10^{10}M_\odot$ can reasonably account for the
two states of S5 0014+813 observed in 1993 (Bechtold et al. 1994;
grey symbols and line) and in Jan 2007 (red symbols).
A still smaller accretion rate can (approximately) account for the 2MASS data.
}
\label{f2}
\end{figure}

In Fig. \ref{f1} we show two theoretical models,
one accounting for the BAT data, but overproducing the $\sim$10 keV flux,
the other accounting for the entire XRT and {\it XMM--Newton} data range
but under--producing the BAT flux.
The two models, bracketing the XRT and the BAT states,
are obtained with a minimal change of the values of the input parameters
(listed in Tab. \ref{para}).
The little differences have no impact on our conclusions.
We used the model discussed in detail in Ghisellini \& Tavecchio (2009),
accounting for the presence of many sources of photons located
externally to the jet.
We would like to stress that the use of a particular jet model
is not crucial for our discussion.
What is important is that the optical--UV emission is dominated by the 
accretion disk emission.
Indeed, as Fig. \ref{f1} shows, the non--thermal emission from the jet
contributes only for a few per cent of the total flux.

\begin{table*} 
\centering
\begin{tabular}{llllllllllll|llll}
\hline
\hline
$R_{\rm diss}$ &$M$ &$R_{\rm BLR}$ &$P^\prime_{\rm i}$ &$L_{\rm d}$ &$B$ &$\Gamma$ &$\theta_{\rm v}$
    &$\gamma_{\rm b}$ &$\gamma_{\rm max}$ &$s_1$  &$s_2$ 
&$\log P_{\rm r}$ &$\log P_{\rm B}$ &$\log P_{\rm e}$ &$\log P_{\rm p}$ \\ 
~[1]      &[2] &[3] &[4] &[5] &[6] &[7] &[8] &[9] &[10] &[11] &[12]  &[13] &[14] &[15] &[16] \\
\hline   
9.6e3 (800) &4e10  &4.9e3 &0.1  &2.4e3 (0.4) &0.22  &16   &3   &70   &3e3   &--1   &2.3 &46.3 &46.6 &45.7 &47.4  \\ 
4.8e3 (400) &4e10  &4.9e3 &0.1  &2.4e3 (0.4) &1.54  &11   &3   &60   &3e3   &--1   &3.6 &46.1 &47.4 &44.5 &47.1 \\ 
\hline
\hline 
\end{tabular}
\caption{Parameters used to model the SED.
Col. [1]: dissipation distance in units of $10^{15}$ cm and (in parenthesis) in units of $R_{\rm S}$;
Col. [2]: black hole mass in solar masses;
Col. [3]: size of the BLR in units of $10^{15}$ cm;
Col. [4]: power injected in the blob (in the comoving frame), in units of $10^{45}$ erg s$^{-1}$; 
Col. [5]: accretion disk luminosity in units of $10^{45}$ erg s$^{-1}$ and
        (in parenthesis) in units of $L_{\rm Edd}$;
Col. [6]: magnetic field in Gauss;
Col. [7]: bulk Lorentz factor at $R_{\rm diss}$;
Col. [8]: viewing angle in degrees;
Col. [9] and [10]: break and maximum random Lorentz factors of the injected electrons;
Col. [11] and [12]: slopes of the injected electron distribution [$Q(\gamma)$] below and above $\gamma_{\rm b}$;
Col. [13]; [14]; [15] and [16]: jet power in the form of radiation, Poynting flux,
bulk motion of electrons and protons (assuming one proton
per emitting electron). These powers are derived quantities, not input parameters.
For a detailed description of the parameters and the model see Ghisellini \& Tavecchio (2009).
}
\label{para}
\end{table*}

\section{Discussion}

\subsection{Comparisons with other black hole mass estimates}

The value of $4\times 10^{10}M_\odot$ for the black hole of S5 0014+813
is unprecedented for a radio--loud source.
Among quasars studied with the velocity width and
continuum luminosity method, there seems to a be a saturation value
around $5\times 10^9$--$10^{10} M_\odot$ (Shen et al., 2008),
with very few black hole masses above $10^{10}M_\odot$
(see also Vestergaard \& Osmer 2009; Vestergaard et al. 2008; 
Kelly, Vestergaard \& Fan 2008; Natarajan \& Treister 2009).

For one specific quasar, Q0105--2634, Dietrich \& Hamann (2004) 
estimated a black hole mass of  $(41.4\pm12.2)\times 10^{10}M_\odot$
using the H$_{\beta}$ line width and the luminosity/BLR size 
in Kaspi et al. (2000), but for the same object the estimate decreases 
to $(5.2\pm1) \times 10^9 M_\odot$
using the MgII line and the McLure \& Jarvis (2002) BLR size/luminosity
relation.
This testifies the large uncertainties related to this method of estimating
the black hole masses.

Our method is free from this kind of uncertainties, but relies mainly
on three assumptions: i) the emission of the disk, apart from a $\cos\theta$ term,
is isotropic;
ii) a standard (optically thick,
geometrically thin) disk and 
iii) a black--body emission at each radius.
While the latter assumption is conservative (since the black--body is the
best radiator, it is bound to give a lower limit to the derived masses and 
accretion rates), the first two hypotheses can seriously affect
our mass determination.
We will first discuss the implications of having found such a large mass,
assuming that 40 billions solar masses is the real value, and then we
will discuss how a non--standard disk can impact our mass estimate.

\subsection{Consequences of the huge black hole mass estimate}

The fact to have found such a huge black hole mass in a blazar has 
a simple and  profound implication: since we selected it on the basis 
of the beamed non--thermal hard X--ray continuum, there must be many 
other sources like S5 0014+813 that are pointing in other 
directions, with a much fainter (de--beamed) non--thermal emission.
The relative number scales as $\Gamma^2\gsim 100$, where $\Gamma$ is the bulk Lorentz
factor of the jet.
These misaligned sources are unnoticeable in the X--ray and $\gamma$--ray bands,
but since the accretion disk is unbeamed (and {\it if} its emission is isotropic), 
they should be detectable in optical all sky surveys, like the SDSS.
Assuming that there are 100 sources as optically bright as  S5 0014+813
in the 30,000 deg$^{-2}$ of the sky excluding the Galactic plane with $|b|<15^\circ$,
the surface density of these objects is of the order of 
$\Sigma\sim 3.3\times 10^{-3}(\Gamma/10)^2$ deg$^{-2}$.
We can estimate how many of these sources the SDSS can detect.   
For the part of the sky already monitored and covered by spectroscopy
by the SDSS (i.e. 5,700 square degrees for quasars) 
we expect $\sim$19 sources like S5 0014+813. 
Having a few objects that are indeed that luminous, 
the SDSS results are in this respect borderline (e.g. Vestergaard 2009). 


We can also compare these estimates with the expectations
of different models/correlations relating the black hole mass (MBH) 
with the velocity dispersion $\sigma$, 
the mass of the host galaxy bulge and its dark mass halo.

Assume first that the relations among these quantities are
redshift--independent.
Then, from Tremaine et al. (2002), a MBH of 40 billion $M_\odot$
should correspond to $\sigma=824$ km s$^{-1}$, and according 
to Ferrarese (2002) this yields $M_{\rm halo}=6.7 \times 10^{13} M_\odot$.
Adopting a Press \& Schechter law we derive 
$\Sigma\sim 0.07$ deg$^{-2}$.

Alternatively, we can derive the mass of the host bulge by applying 
the relation  $M_{\rm BH,8}\sim 1.68 M_{\rm bulge,11}^{1.12}$
proposed by Haring \& Rix (2004), finding $M_{\rm bulge}=1.4 \times 10^{13}M_\odot$. 
This $M_{\rm bulge}$ can be related to $\sigma$ by assuming 
the ``fundamental plane of black holes" proposed by Hopkins et al. (2007),
finding $\sigma\sim10^3$ km s$^{-1}$, and so $M_{\rm halo}=1.3\times 10^{14}M_\odot$,
corresponding to $\Sigma\sim 7\times 10^{-4}$ deg$^{-2}$.

The above estimates assumed that the correlations between black holes 
and their host are redshift--independent.
However, the 
$M_{BH}$--$M_{bulge}$ and the $M_{BH}$--$\sigma$ correlations
might evolve with the cosmic time, as suggested by 
McLure et al. (2006); Peng et al. (2004); Treu et al. (2004); Woo et al. (2006). 
Mc Lure et al. (2006) suggested that the ratio
$M_{BH} /M_{\rm bulge}$ can evolve as $(1+z)^2$.
In this case, $M_{\rm bulge}\sim 2.3\times 10^{12} M_\odot$.
Through the fundamental plane we find 
$\sigma\sim$1800 km s$^{-1}$ and  $M_{\rm halo}=5.1\times 10^{14}M_\odot$.
This large halo mass makes the surface density to drop to
$\Sigma\sim 2\times 10^{-9}$ deg$^{-2}$.
Note however that such a high velocity dispersion is rather extreme, 
implying a very dense bulge, with a size of $\sim 3$ kpc, within a very large halo.

Finally, Woo et al. (2008) proposed that, at a fixed velocity dispersion,
the MBH scales with redshift as $(1+z)^3$. 
This means that, at large redshifts, we can find larger MBH
within hosts with smaller bulge and halo masses.
We then use the Tremaine et al. (2002) relation to find 
$\sigma=257$ km s$^{-1}$, corresponding to 
$M_{\rm halo}=3.5 \times 10^{12} M_\odot$.
This relatively small halo mass corresponds to a very large
surface density: $\Sigma\sim 6000$ deg$^{-2}$.

We can conclude that the current predictions for the
number density of the largest MBH are far from conclusive, essentially because
we are far in the exponential tail of the distributions, and small changes
of the host properties can dramatically change the predicted numbers:
while the first two estimates of $\Sigma$ are not far from the limits derived 
from our (single) object, the last two are either too small or too large.


\subsection{Super--Eddington disks}

The observed large optical luminosity is at the base of our black
hole mass estimate. 
While we can exclude that it is dominated by beamed jet emission,
we discuss here the possibility that the underlying accretion
disk producing it might by non--standard.
In the literature, two main alternatives have already been proposed:
geometrically thick and radiation supported disk, for super--Eddington
accretion rates (Jaroszynski, Abramowicz \& Paczynski 1980) and the so--called slim disk, 
with accretion rates close to the Eddington one (Abramowicz et al. 1988).
The latter can emit a super--Eddington luminosity 
because the advection of the flow helps gravity to sustain the
radiation force, but are characterised by a relatively small
height to radial distance ratio ($H/R$) that does not allow a strong collimation
of the produced radiation.
The foreseen emission is not a pure black--body, but a a modified one, since
electron scattering is very important.
This implies larger temperatures than in the standard case 
(Szuszkiewicz et al. 1996),
that might be a problem in our case, due to the relatively severe
upper limits in the UV.

In the thick disk case, instead, the emission is locally close to Eddington,
but the presence of a narrow funnel can collimate (via electron scattering)
the radiation produced deep in the inner funnel into a narrow cone.
Observing at small angles from the axis of the funnel, we can then
have the impression of a super--Eddington luminosity (up to a factor of 10--20).
However, as detailed in Madau (1988), it is the high
frequency radiation that is boosted the most (being the one produced mostly in the
inner funnel), and this can be
a problem in our case.

\section{Conclusions}

The optical--UV luminosity of the blazar S5 0014+813 exceeds
$10^{48}$ erg s$^{-1}$ and through the construction of the 
overall SED it can be convincingly associated to the radiation
produced by an accretion disk.
The found black hole mass is 40 billion of solar masses accreting
at the 40\% of the Eddington rate, and in the past it might 
have reached the full Eddington rate.
Since this source was found because of its relativistically beamed hard X--ray emission,
there should be many other sources of same mass and accretion rate,
but whose jet is pointing in other directions.
Current theoretical estimates do not exclude this, but are very uncertain.

There are ways to reduce the estimated black hole mass, invoking non--standard
accretion disks, that however tend to emit a spectrum bluer than observed.
On the other hand, the real physical properties of these slim or thick
disks may be somewhat different from the assumed ones, and we cannot
rule them out. 
A more definite mass estimate would also greatly benefit by simultaneous data,
from the far IR to the UV.

The fact that this very large black hole mass has been found in a radio--loud source
may not be a coincidence, if the presence of a jet is a crucial ingredient for the 
transfer of the angular momentum of the accreting matter, allowing the black hole
to grow faster, as pointed out by Jolley \& Kuncic (2008).
In this case the found black hole mass can be real, or else
the presence of the jet induces a super--Eddington accretion rate, 
making the disk slim or helping the formation of a funnel.
Either way, S5 0014+813 is an exceptional source, worth to be
investigated further.

\section*{Acknowledgments}
We thank the referee, E. Pian, for useful suggestions and
criticism.
This work was partly financially supported by a 2007 COFIN-MIUR 
and an ASI I/088/06/0) grants.
This research has made use of the NASA/IPAC Extragalactic Database (NED) 
which is operated by the Jet Propulsion Laboratory, California Institute 
of Technology, under contract with NASA and
of data obtained from the High Energy Astrophysics Science Archive Research Center 
(HEASARC), provided by NASA's Goddard Space Flight Center.

\end{document}